\documentclass[prd,aps,twocolumn,a4paper,nofootinbib]{revtex4-1}

\newif\ifusesec
\usesectrue 
   
\usepackage{graphicx,psfrag}
\usepackage{mathrsfs}
\usepackage{amsmath,amsfonts,amssymb}
\usepackage{multirow}
\usepackage{comment}

\DeclareSymbolFontAlphabet{\mathrsfs}{rsfs}
\DeclareMathAlphabet\mathbfcal{OMS}{cmsy}{b}{n}

\newcommand{\be}{\begin{equation}}
\newcommand{\ee}{\end{equation}}
\newcommand{\bea}{\begin{eqnarray}}
\newcommand{\eea}{\end{eqnarray}}
\newcommand{\bel}{\begin{align}}
\newcommand{\eel}{\end{align}}

\def\GMc2{G M_{\odot} c^{-2}}

\DeclareSymbolFontAlphabet{\mathrsfs}{rsfs}
\DeclareMathAlphabet{\mathcal}{OMS}{cmsy}{m}{n}

\usepackage{color}

\definecolor{cyan}{rgb}{0,0.9,0.9}
\definecolor{orange}{rgb}{0.9,0.5,0}
\definecolor{magenta}{rgb}{1,0,1}
\definecolor{purple}{rgb}{0.8,0.4,0.8}
\definecolor{gray}{rgb}{0.8242,0.8242,0.8242}

\begin{document}

\title{Analytical determination of the two-body gravitational interaction potential \\ at the 4th post-Newtonian approximation}

\author{Donato \surname{Bini}$^1$}
\author{Thibault \surname{Damour}$^2$}

\affiliation{$^1$Istituto per le Applicazioni del Calcolo ``M. Picone'', CNR, I-00185 Rome, Italy}
\affiliation{$^2$Institut des Hautes Etudes Scientifiques, 91440 Bures-sur-Yvette, France}

\date{\today}

\begin{abstract}
We complete the analytical determination, at the 4th post-Newtonian approximation, of the main radial potential describing  the gravitational interaction of two bodies within the effective one-body formalism. The (non logarithmic) coefficient $a_5 (\nu)$ measuring this 4th post-Newtonian interaction potential is found to be linear in the symmetric mass ratio $\nu$. Its $\nu$-independent part $a_5 (0)$ is obtained by an analytical gravitational self-force calculation that unambiguously resolves the formal infrared divergencies which currently impede its direct post-Newtonian calculation. Its $\nu$-linear part $a_5 (\nu) - a_5 (0)$ is deduced from recent results of Jaranowski and Sch\"afer, and is found to be significantly negative.
\end{abstract}


\maketitle

\section{Introduction}
\label{sec:intro}

A ground-based network of interferometric gravitational wave detectors is currently being upgraded, and is expected to detect, in the near future, the gravitational wave signals emitted during the late inspiral and merger of compact binaries. The detection and data analysis of these signals require very accurate theoretical predictions of the motion of compact binaries and its associated gravitational wave emission. It has become clear over the past few years that the best way to meet the latter theoretical challenge is to combine knowledge coming from various techniques: post-Newtonian (PN) expansions, post-Minkowskian ones, black-hole perturbation theory, the effective-one-body (EOB) formalism, gravitational self-force (GSF) calculations, and full numerical simulations.

In this paper we report the first analytical determination of the 4th post-Newtonian (4PN) contribution to the radial interaction potential $A(r;m_1,m_2)$ of a general relativistic two-body system (with masses $m_1$ and $m_2$). The interaction potential $A(r;m_1,m_2)$ is a {\it gauge-invariant} function which enters the EOB formalism \cite{Buonanno:1998gg,Buonanno:2000ef,Damour:2000we,Damour:2001tu}. It is a useful generalization of the well-known Schwarzschild potential $A^S (r) = 1-2GM/c^2r$. The EOB formalism maps the {\it conservative} dynamics of a (non spinning) two-body system $(m_1,m_2)$ onto the geodesic dynamics of one body of mass $\mu = m_1m_2 / (m_1 + m_2)$ in a stationary and spherically symmetric ``effective'' metric,
\begin{eqnarray}
\label{eq1}
ds_{\rm eff}^2 &= &-A (r;m_1,m_2) c^2 dt^2 \nonumber \\
&+ &B(r;m_1,m_2) dr^2 + r^2 (d\theta^2 + \sin^2 \theta d\varphi^2) \, ,
\end{eqnarray}
together with post-geodesic corrections described by a function $Q(r,p_r , p_{\varphi} ; m_1 , m_2)$ which is, at least, quartic in the (EOB) radial momentum $p_r$ \cite{Damour:2000we}. The gauge-invariant dynamics of the sequence of circular orbits is fully encoded in the sole radial potential $A(r;m_1,m_2)$ \cite{Damour:2009sm,Barausse:2011dq,Akcay:2012ea}.

Introducing the notation $M := m_1 + m_2$, $\nu := \mu / M = m_1 m_2 / (m_1 + m_2)^2$ and $u := GM/c^2r$, the PN expansion, up to the 4PN level included, of the radial potential $A(r;m_1,m_2) \equiv A(u;\nu)$ has the form
\begin{eqnarray}
\label{eq2}
A(u;\nu) &= &1-2u + \nu a_3(\nu) u^3 + \nu a_4 (\nu) u^4 \nonumber \\
&+ &\nu (a_5^c (\nu) + a_5^{\ln} (\nu) \ln u) u^5 + o(u^5) \, .
\end{eqnarray}
When the symmetric mass ratio $\nu = \mu / M$ tends toward zero $A(u;\nu)$ reduces to the Schwarzschild potential $A^S (u) = A(u;0)=1-2u$. Each additional term $\sim \nu a_n (\nu) u^n$ represents the contribution of the $(n-1)$-th PN approximation.  The values of the 1PN and 2PN coefficients (namely  $a_2 (\nu) = 0$ and $a_3 (\nu) = 2$) were derived in \cite{Buonanno:1998gg} from the 2PN Delaunay Hamiltonian of \cite{Damour:1988mr}. The value of the 3PN-level coefficient,
\be
\label{eq4}
a_4 (\nu) = \frac{94}3 - \frac{41}{32} \, \pi^2 \, ,  
\ee
was derived in \cite{Damour:2000we} from the 3PN Hamiltonian of \cite{Jaranowski:1997ky,Damour:2001bu}. The value of the 4PN-level {\it logarithmic} coefficient ,  
\be
\label{eq5}
a_5^{\ln} (\nu) = \frac{64}5 \, ,
\ee
was derived in \cite{Damour:2009sm,Blanchet:2010zd,Barausse:2011dq} from the results of Ref.~\cite{Blanchet:1987wq}.

In this work we shall derive, for the first time, the analytical value of the non-logarithmic 4PN-level coefficient in Eq.~(\ref{eq2}). We find (with $\gamma$ denoting Euler's constant)
\begin{subequations}  \label{eqs6}
\begin{eqnarray}
\label{eq6a}
a_5^c (\nu) &= &a_5^{c0} + \nu a_5^{c1}  \, ,\\
\label{eq6b}
a_5^{c0} &=  &-\frac{4237}{60} + \frac{2275}{512} \pi^2 + \frac{256}5 \ln 2 + \frac{128}5 \gamma \, ,\\
\label{eq6c}
a_5^{c1} &= &-\frac{221}6 + \frac{41}{32} \pi^2 \, .
\end{eqnarray}
\end{subequations}
We shall discuss below the compatibility of our analytic result Eq.~(\ref{eq6b}) with the numerical estimates of $a_5^{c0}$ which have been recently inferred \cite{Tiec:2011ab, Tiec:2011dp,Barausse:2011dq} from  accurate numerical computations of Detweiler's redshift function \cite{Detweiler:2008ft,Blanchet:2009sd,Blanchet:2010zd}. 

We obtained the above results by combining several different methods. The $\nu$-dependent part of $a_5^c (\nu)$ was derived from the recent computation by Jaranowski and Sch\"afer \cite{Jaranowski:2013lca} of the 4PN contributions to the two-body Hamiltonian coming from {\it infra-red-convergent} near-zone effects. By contrast, the $\nu \to 0$ part of $a_5^c (\nu)$ (i.e. the coefficient $a_5^{c0}$, Eq.~(\ref{eq6b})) is  connected with {\it formally infra-red-divergent} effects that could not be controlled in \cite{Jaranowski:2013lca}. More precisely, the physics behind the value of $a_5^{c0}$ is (partly) linked to the tail-transported, {\it hereditary} influence of the past evolution of the binary system on its present dynamics. This hereditary influence was elucidated fifteen years ago by Blanchet and Damour \cite{Blanchet:1987wq}, and was indeed shown to enter the dynamics at the 4PN level, and to signal a breakdown of the separation between near-zone and wave-zone effects. As a consequence, our derivation of the value of $a_5^{c0}$ had to go beyond the usual PN method by incorporating the transition between near-zone and wave-zone physics. This was done by a combination of techniques. First, we use a recently discovered link \cite{Tiec:2011ab,Tiec:2011dp,Barausse:2011dq} between the $O(\nu)$ piece of the EOB (gauge-invariant) radial potential $A(u;\nu)$ and the $O(\nu)$ piece of Detweiler's gauge-invariant ``redshift'' function $z_1 (\Omega)$ \cite{Detweiler:2008ft}, along circular orbits (of orbital frequency $\Omega$). Second, we used a combination of GSF techniques for {\it analytically} computing the $O(\nu)$ piece of $z_1 (\Omega)$, namely: spherical-harmonics-mode-sum regularization \cite{Barack:1999wf,Detweiler:2008ft}, and improved analytic black hole perturbation techniques developed by the Japanese relativity school \cite{Mano:1996vt,Mano:1996mf,Mano:1996gn,Sago:2002fe,Nakano:2003he,Hikida:2004jw,Hikida:2004hs}.

\section{Analytical computation of conservative GSF effects along circular orbits}
\label{sec:two}

Detweiler \cite{Detweiler:2008ft} has emphasized the existence of one conservative, gauge-invariant function, available within first-order $(O(\nu))$ GSF theory, associated with the sequence of {\it circular} orbits of an extreme mass-ratio binary system: $m_1 \ll m_2$. Computing the $O(\nu)$ piece of this redshift function $z_1 (\Omega ; \nu) \equiv 1/u^t_1 (\Omega;\nu)$ is equivalent \cite{Detweiler:2008ft,Blanchet:2009sd,Akcay:2012ea} to computing the {\it regularized} value, along the worldline $y_1^{\mu}$ of the small mass $m_1$, of the double contraction of the $O(m_1)$ metric perturbation $h_{\mu\nu}$ (considered in an {\it asymptotically flat} gauge),
\be
\label{eq18}
g_{\mu\nu} (x;m_1,m_2) = g_{\mu\nu}^{(0)} (x;m_2) + m_1 h_{\mu\nu} (x) + O(m_1^2) 
\ee
(where $g_{\mu\nu}^{(0)} (x;m_2) \!= \!g_{\mu\nu} (x;m_1 \!= \!0 , m_2)$ is a Schwarzschild metric of mass $m_2$), with the four-velocity $u_1^{\mu} = dy_1^{\mu} / ds_1$ of $m_1$, say
\be
\label{eq18bis}
h_{uu}^R := {\rm Reg}_{x \to y_1} [h_{\mu\nu} (x) u_1^{\mu} u_1^{\nu}] \, .
\ee
Following Refs.~\cite{Barack:1999wf,Detweiler:2002gi,Hikida:2004jw,Detweiler:2008ft,Blanchet:2009sd} the regularization operation indicated in Eq.~(\ref{eq18bis}) is done by subtracting the (leading-order) singular part in the spherical harmonics expansion of $h_{uu}$. This yields $h_{uu}^R$ as a series indexed by $l=0,1,2,\ldots$: $h_{uu}^R = \sum_{l=0}^{\infty} (h_{uu}^{(l)} - D_0)$, where $h_{uu}^{(l)} = \sum_{m=-l}^{+l} u^{\mu} u^{\nu} h_{\mu\nu}^{(l,m)}$, and where the ($l$-independent) subtraction constant $D_0$ is known \cite{Detweiler:2002gi,Detweiler:2008ft} to be
\begin{eqnarray}
\label{eq21}
D_0 &= &2u\sqrt{\frac{1-3u}{1-2u}} F \left( \frac12 , \frac12 , 1 ; \frac{u}{1-2u} \right)  
\end{eqnarray}
Here, $u = Gm_2 / c^2 r = GM/c^2 r + O(\nu)$, and $F(a,b,c;z)$ is Gauss's hypergeometric function.

In the following, we find more convenient to work with the double contraction $h_{kk} := h_{\mu\nu} k^{\mu} k^{\nu}$ with the helical Killing vector $k^{\mu} \partial_{\mu} = \partial_t + \Omega \partial_{\varphi}$ (such that $k^{\mu} = (ds_1 / dt) u_1^{\mu}$ with $ds_1 / dt = \sqrt{1-3u}$), i.e.
\be
\label{eq22}
h_{kk}^R = (1-3u) h_{uu}^R = \sum_l \left( \left( \sum_m h_{kk}^{(l,m)} \right) - \tilde D_0 \right) ,
\ee
with a renormalized subtraction constant $\tilde D_0 := (1-3u) D_0$.

In previous works \cite{Detweiler:2008ft,Sago:2008id,Blanchet:2009sd,Akcay:2012ea}, $h_{uu}^R$ was evaluated numerically along a (discrete) sequence of circular orbits [parametrized by $r$, $u$ or $x=(GM\Omega / c^3)^{2/3} = u + O(\nu)$].  So far, the analytical knowledge of   $h_{uu}^R$  was limited to the 3PN level \cite{Blanchet:2009sd}, except for the logarithmic contributions at the 4PN \cite{Damour:2009sm} and 5PN \cite{Blanchet:2010zd} levels.
Here, we report, for the first time, on a complete {\it analytical} computation of  $h_{kk}^R = (1-3u) h_{uu}^R$ at the 4PN level, i.e. up to the fifth power of the gravitational potential $u = GM/c^2 r$. We computed it essentially by considering the {\it post-Minkowskian} expansion of the function $h_{kk}^R (u)$ (weak-field expansion in powers of $G$), without being limited by the breakdown of the usual post-Newtonian expansion arising at the 4PN level \cite{Blanchet:1987wq}. The theoretical tools for computing the post-Minkowskian expansion of Regge-Wheeler-Zerilli (RWZ) black hole perturbation theory have been developed by Mano, Suzuki and Takasugi \cite{Mano:1996vt,Mano:1996mf,Mano:1996gn}. [Previous valient attempts to apply the latter formalism to GSF theory were bogged down by gauge-dependent issues \cite{Sago:2002fe,Nakano:2003he,Hikida:2004hs}.] We summarize here the main features of our analysis, leaving details to a future exposition.

In the series (\ref{eq22}) over $l$, the low-multipole contributions $l=0$ and $l=1$ (even and odd) can be computed from the corresponding exact results of Zerilli \cite{Zerilli:1971wd}. On the other hand, the ``dynamical'' multipoles of order $l \geq 2$ are more difficult to evaluate. We started from the {\it corrected} version of the RWZ equations derived by Sago, Nakano and Sasaki \cite{Sago:2002fe,Nakano:2003he,Hikida:2004jw}.
The  RWZ formalism expresses an {\it odd-parity} metric perturbation $h_{\mu\nu}^{(l,m)}$, with frequency $\omega$, in terms of a radial function $R_{lm\omega}^{({\rm odd})} (r)$ satisfying a Regge-Wheeler (RW)-type equation
\be
\label{eq23}
{\mathcal L}^{(r)}_{\rm (RW)}[R_{lm\omega }^{\rm (odd)}]=S_{lm\omega}^{\rm (odd)}(r) \, .
\ee
Here ${\mathcal L}^{(r)}_{\rm (RW)}$ denotes the RW operator
\be
\label{eq24}
{\mathcal L}^{(r)}_{\rm (RW)}=f^2(r) \frac{d^2}{dr^2} +\frac{2M}{r^2} f(r) \frac{d}{dr }  + [\omega^2 -V_{\rm (RW)}(r)] \, , 
\ee
with a RW potential $V_{\rm (RW)}(r)=f (r) \left(\frac{L}{r^2}-\frac{6M}{r^3}  \right)$; 
where $L:=l(l+1)$, and $f(r) := 1-2M/r$. [Here, as often in the following, we omit to include a label $l\omega$ indicating the $l\omega$-dependence of various objects.] 
The odd-parity source term in Eq.~(\ref{eq23}) (given by Eq.~(A35) of \cite{Sago:2002fe}) is of the form
\be
\label{eq26}
S_{lm\omega}^{\rm (odd)}(r)=s^{\rm (o)}_0\delta(r-r_0)+s^{\rm (o)}_1\delta'(r-r_0)\,,
\ee
where $r_0$ denotes the radius of the circular orbit of particle 1. On the other hand, the original RWZ formalism expresses a (monochromatic) {\it even-parity} $h_{\mu\nu}^{(l,m)}$ in terms of a radial function $Z_{lm\omega} (r)$ satisfying a Zerilli-type equation \cite{Zerilli:1971wd}, involving a more complicated potential than the RW equation (\ref{eq23}). Using a result of Chandrasekhar \cite{Chandrasekhar:1975}, one can, however, replace the pair ${\mathcal Z}_{lm\omega} := (Z_{lm\omega} (r) , dZ_{lm\omega} (r)/dr)$ by a new pair of functions, say ${\mathcal R}_{lm\omega} := (R_{lm\omega}^{({\rm even})} (r) , dR_{lm\omega}^{({\rm even})} (r)/dr)$, satisfying a {\it simpler RW-type} equation, say
\be
\label{eq27}
{\mathcal L}^{(r)}_{\rm (RW)}[R_{lm\omega}^{\rm (even)}]=S_{lm\omega}^{\rm (even)}(r)\,.
\ee

The price to pay for this simplification of the potential is: (i) the transformation between ${\mathcal Z}_{lm\omega}$ and ${\mathcal R}_{lm\omega}$ involves source terms, and (ii) the new even-parity source term is connected to the original Zerilli one by an expression of the form $S_{lm\omega}^{\rm (even)}={\mathcal A}_{11} (r) S_{lm\omega}^{\rm (Z)}+{\mathcal A}_{12}(r) \frac{d}{dr} (S_{lm\omega}^{\rm (Z)})$. As a consequence the new even-parity source term in Eq.~(\ref{eq27}) is of the form
\be
\label{eq28}
S_{lm\omega}^{\rm (even)} =  s^{\rm (e)}_0\delta(r-r_0)+s^{\rm (e)}_1\delta'(r-r_0)+s^{\rm (e)}_2\delta''(r-r_0) \, .
\ee

At this stage, the problem is reduced to solving some RW equation (one for each $lm\omega$ and each parity) with given (distributional) source terms. The source terms derive from the spherical harmonics projection of the distributional stress-energy tensor of particle 1: $
T_1^{\mu\nu} (x^{\lambda}) = m_1 (-g)^{-1/2} \int ds_1 u_1^{\mu} u_1^{\nu} \delta^{(4)} (x^{\lambda} - y_1^{\lambda} (s_1)) $.
As a consequence, in the case of a circular orbit, the (discrete) frequencies entering $T_1^{\mu\nu} (x)$ (and therefore $h_{\mu\nu} (x)$) are related to the basic orbital frequency $\Omega = d\phi_1 / dt_1$ and the ``magnetic'' number $m$ by $\omega = \omega_m := m\Omega$.

The solution of the RW equations (\ref{eq23}), (\ref{eq27}) is determined by the choice of the boundary conditions incorporated in a Green function, normalized so as to satisfy ${\mathcal L}^{(r)}_{\rm (RW)} G(r,r')=f(r')\delta (r-r')$.
Such a Green function can be expressed in terms of two, specially chosen, independent homogeneous solutions of the RW operator, and of the Heaviside step function $H(x)$:
\begin{eqnarray}
\label{eq31}
G(r,r')&=&\frac{1}{W}\Bigl[X_{\rm (in)}(r)X_{\rm (up)}(r')H(r'-r) \nonumber \\
&+&X_{\rm (in)}(r')X_{\rm (up)}(r)H(r-r')  \Bigl] \, .
\end{eqnarray}
Here $W$ denotes the (constant) Wronskian $W=f(r) [X_{\rm (in)}(r) \, d X_{\rm (up)}(r )/dr
-dX_{\rm (in)}(r)/dr X_{\rm (up)}(r) ]$.
The physical Green function we are interested in is the {\it retarded} one. It is obtained, as usual, by choosing for $X_{({\rm in})}^{l\omega}$ a solution of  ${\mathcal L}_{({\rm RW})}^{(r)} X_{({\rm in})} (r) = 0$ that is incoming from $r = +\infty$ (and purely ingoing on the horizon), and for $X_{({\rm up})}^{l\omega}$ a solution of $ {\mathcal L}_{({\rm RW})}^{(r)} X_{({\rm up})} (r)=0$  that is upgoing from the horizon (and purely outgoing at infinity). This uniquely determines the solutions of the even-parity and odd-parity RW equations, namely
\be
\label{eq33}
R_{lm\omega}^{\rm (even/odd)}(r)=\int dr' G(r,r')f(r')^{-1}S_{lm\omega}^{\rm (even/odd)}(r')\,.
\ee

Note that the distributional nature of the radial source functions, notably $S_{lm\omega}^{\rm even} (r) \ni \delta'' (r-r_0)$, implies that, e.g., $R_{lm\omega}^{\rm even} (r)$ is not only discontinuous as $r$ crosses $r_0$, but (formally) contains a contribution $\propto \delta (r-r_0)$.

Having determined $R_{lm\omega}^{({\rm even}/{\rm odd})} (r)$ by the (distributional) formula (\ref{eq33}), one can then compute the original Zerilli radial functions $(Z_{lm\omega} (r) , dZ_{lm\omega} (r) / dr)$, and thereby evaluate the metric perturbation $h_{\mu\nu}^{(lm\omega)} (r)$. The next step is to consider $h_{kk}(t,r,\theta,\phi)$ at field-point values of
 $t, \theta$ and $\phi$  corresponding to the considered instantaneous position of particle~1, say $t,\theta = \pi / 2$ and $\phi = \phi_1 (t) = \Omega t$ (in the equatorial plane of the background Schwarzschild metric).  At this stage, $h_{kk}$ depends only on $r$. Considering the two limits $r \to r_0^-$ and $r \to r_0^+$, we have checked that they yield the same result for the value of the gauge-invariant quantity $h_{kk}$ at the location $r=r_0$ of particle~1.  [This confirms the idea of Detweiler \cite{Detweiler:2008ft} that the gauge-invariant quantity $h_{uu}^R$ can be correctly evaluated on the worldline of $y_1$ even if one uses a gauge (such as the RWZ one) where $h_{\mu\nu} (x;y_1)$ has a worse behaviour than its Lorenz-gauge version.]

Our final result for $h_{kk}^{(l,m)} \equiv h_{kk}^{(l,m)} (r_0)$, which enters Eq.~(\ref{eq22}), is the sum of an even and an odd contribution, where the latter one takes the rather simple form
\begin{eqnarray}
\label{eq35}
h_{kk,lm}^{\rm (odd)}&=&-|\partial_\theta Y_{lm}(\pi/2,0)|^2 \frac{8 \pi u_1^t}{r_0^3W\Lambda} M f_0^2 \nonumber \\
&\times&\left[r_0 \frac{d X_{l\omega}^{\rm (in)}}{dr_0}+X_{l\omega}^{\rm (in)}\right] 
\left[r_0 \frac{d X_{l\omega}^{\rm (up)}}{dr_0}+X_{l\omega}^{\rm (up)}\right]\,. 
\end{eqnarray}
Here $ u_1^t = (1-3u)^{-1/2}$, and $\Lambda := \frac14 (l-1) l(l+1)(l+2)$. The corresponding result for $h_{kk,lm}^{({\rm even})}$ has a similar structure $\propto - ({\rm source})^2 \times {\mathcal F} (X_{l\omega}^{({\rm in})}) {\mathcal F} (X_{l\omega}^{({\rm up})})$, with a squared-source term $\propto \vert Y_{lm} (\pi/2,0)\vert^2$ and a product of two identically constructed combinations of $X$ and $dX/dr_0$, evaluated for $X_{l\omega}^{({\rm in})}$ and $X_{l\omega}^{({\rm up})}$. [These expressions have the usual ``one-loop'' structure $({\rm source}) \times (\mbox{Green function}) \times ({\rm source})$.]

To evaluate $h_{kk}^R$ from the RWZ result (\ref{eq35}) (and its even-parity analog) one still needs, according to Eq.~(\ref{eq22}), to: (i) sum over $m$ from  $-l$ to $+l$; (ii) subtract $\tilde D_0$; and, finally, (iii) sum the result of (i) and (ii) over $ l = 0,1,2,\ldots$. Even the first (finite) sum over $m$ is quite nontrivial to compute analytically because one must remember that the index $\omega$ on the two solutions $X_{l\omega}^{({\rm in})} , X_{l\omega}^{({\rm up})}$ entering (\ref{eq35}) (and its even-parity analog) actually refers to $\omega_m = m\Omega$, not to mention the fact that one needs to obtain explicit, analytic expressions for the two homogeneous solutions $X_{l\omega}^{({\rm in})} (r)$ and $X_{l\omega}^{({\rm up})} (r)$. The latter problem has been formally solved by Mano et al. \cite{Mano:1996vt,Mano:1996mf,Mano:1996gn} who gave analytic expressions for $X_{l\omega}^{({\rm in})}$ and $X_{l\omega}^{({\rm up})}$ in the form of series of hypergeometric functions (of the usual, Gauss, type of $X^{({\rm in})}$ and of the confluent type for $X^{({\rm up})}$). 

It would be quite difficult to use the Mano-type hypergeometric series to compute $h_{kk}^{(l,m)} (r_0)$ for all values of $l$ and $m$. However, the work of Ref.~\cite{Blanchet:1987wq} has shown that, at the 4PN level, the subtle (formally infra-red divergent) mixing of near-zone and wave-zone effects only occurs through quadrupolar $(l=2)$ couplings. This indicates that the full power of the hypergeometric series expansions  is only needed to correctly get the $l=2$ contribution to $h_{kk}^{(l,m)}$, and that a usual PN expansion is accurate enough to evaluate the $l \geq 3$ contributions. We have explicitly checked the correctness of this expectation.  More precisely, we found that the crucial ``beyond-PN'' information is contained in the hypergeometric {\it up} solution  for $l=2$. 
As for the ingoing hypergeometric solution $X_{l\omega}^{({\rm in})} (r)$ we found that, modulo an inessential constant prefactor, it is correctly evaluated by solving the corresponding homogeneous RW equation by a formal PN scheme. Similarly, the solutions $X_{l\omega}^{({\rm in})} , X_{l\omega}^{({\rm up})}$ for $l \geq 3$ can be evaluated with sufficient accuracy by looking for PN-expanded homogeneous solutions of the form (with $\eta :=1/c$)
\begin{eqnarray}
\label{eq41}
X_{l\omega}^{{\rm in} ({\rm PN})} (r) &= &r^{l+1} [1+\eta^2 A_2^{(l)} + \eta^4 A_4^{(l)} \nonumber \\
&+&\eta^6 A_6^{(l)} + \eta^8 A_8^{(l)} + \ldots] \, ,
\end{eqnarray}
\begin{eqnarray}
\label{eq42}
X_{l\omega}^{{\rm up} ({\rm PN})} (r) &= &r^{-l} [1+\eta^2 A_2^{(-l-1)} + \eta^4 A_4^{(-l-1)} \nonumber \\
&+&\eta^6 A_6^{(-l-1)} + \eta^8 A_8^{(-l-1)} + \ldots] \, .
\end{eqnarray}
$A_2^{(l)} , A_4^{(l)} , \ldots$ are certain polynomials in $X_1 = GM/r$, $X_2 = (\omega r)^2$ with $l$-dependent coefficients, modulo some logarithmic corrections $A_{2k}^{(l) \log} \ln (r/R_{2k}^{(l)})$ that must be included in $A_6^{(l)}$ and $A_8^{(l)}$. In addition, the coefficients entering the {\it up} PN solution $X_{l\omega}^{{\rm up} ({\rm PN})}$, Eq.~(\ref{eq42}), are obtained from the $A_{2k}^{(l)}$ coefficients entering the {\it in} PN solution Eq.~(\ref{eq41}) simply by replacing $l$ by $-l-1$ (except for a $4PN$-level term in $A_8^{(l)}$ containing $l+3$ in the denominator which, when changing $l \to -l-1$ and considering the explicit integer value $l=2$, generates a new logarithmic term). We checked that all the arbitrary scales $R_{2k}^{(l)}$, $R_{2k}^{(-l-1)}$ entering the logarithms in these PN expansions drop out of our present 4PN-level computation. 

When inserting  the hypergeometric or PN results for $X_{l\omega}^{({\rm in})} , X_{l\omega}^{({\rm up})}$  in  the expressions (of the type (\ref{eq35})) giving $h_{kk,lm}$ we get explicit results which depend both on $l$ and on $m$ via the $m$-dependent value of $\omega = m\Omega$. The summation over $m$ in Eq.~(\ref{eq22}) then generates finite sums most of which are of the form $
S_{N,l} = \sum_{m=-l}^{+l} m^N \vert Y_{lm} (\pi/2,0)\vert^2
$,
or
$
S'_{N,l} = \sum_{m=-l}^{+l} m^N \vert \partial_{\theta} Y_{lm} (\pi/2,0)\vert^2$.
These sums vanish when the (non negative) integer $N$ is odd, and can be expressed as polynomials in $l$ when $N$ is even, thanks to the results of the Japanese relativity school (see Appendix F in \cite{Nakano:2003he}).
In addition to these sums, our 4PN-accurate calculation of $h_{kk}^R$ involved a new, and more delicate, sum (related to the results of \cite{Blanchet:1987wq}) of the type $
S^{\log}_{N,l} = \sum_{m=-l}^{+l} m^N \ln (-im) \vert Y_{lm} (\pi/2,0)\vert^2$
for $N=6$ and $l=2$. The sum $S_{6,2}^{\log}$ is  real (which is related to the {\it conservative} character of $h_{kk}^R$) and equal to $60 \ln 2 / \pi$.

After  explicitly performing the summation over $m$, and subtracting the $u$-expansion of $\tilde D_0 = (1-3u) D_0$ (with Eq.~(\ref{eq21})), we obtain, according to Eq.~(\ref{eq22}), an explicit expression for $h_{kk}^R$ given by a sum of a few explicit first terms (corresponding to $l = 0,1$ and $2$; with even and odd contributions), plus an infinite series over $l \geq 3$. 
The convergent series entering our calculation are of the form $\sum_{l \geq 3} P_{n-2} (l)/Q_n(l)$ with complicated polynomials of degree $n-2$ and $n$ respectively. They can all be evaluated (after decomposing them in partial fractions in $l$) in terms of $\zeta (2) = \sum_l 1/(l+1)^2 = \pi^2 / 6$. This leads to our final 4PN-accurate result
\begin{eqnarray}
\label{eq46}
h_{kk}^R&=&-2u+5u^2+\frac{5}{4}u^3+\left(-\frac{1261}{24}+\frac{41}{16}\pi^2\right)u^4\nonumber\\
&& +\biggl(\frac{157859}{960}-\frac{256}{5}\gamma-\frac{128}{5}\ln(u)-\frac{512}{5}\ln(2) \nonumber \\
&&-\frac{2275}{256}\pi^2\biggl)u^5 + o(u^5) \, .
\end{eqnarray}

\section{4PN-accurate computation of the EOB radial potential $A(u;\nu)$}
\label{sec:three}


\subsection{4PN-accurate computation of the $O(\nu)$ piece of $A(u;\nu)$}

In this subsection, we consider the ``GSF expansion'' \cite{Damour:2009sm} of the $A(u;\nu)$ radial potential, i.e. its expansion in powers of $\nu$, say
\be
\label{eq47}
A(u;\nu) = 1-2u + \nu a(u) + \nu^2 a_2 (u) + O(\nu^3) \, .
\ee
The work of Refs.~\cite{Tiec:2011ab,Tiec:2011dp,Barausse:2011dq} has led to the following simple relation between the $O(\nu)$ piece $a(u)$ in $A(u;\nu)$ and the $O(m_1/m_2)$ GSF function $h_{kk}^R (u)$, 
\be
\label{eq48}
a(u) = -\frac12 \, h_{kk}^R (u) - \frac{u(1-4u)}{\sqrt{1-3u}} \, .
\ee
Here, it is written in the form used in \cite{Akcay:2012ea} (when using an asymptotically flat gauge $h_{\mu\nu}$, as we are doing here). Note that this relation was used in \cite{Barausse:2011dq} and \cite{Akcay:2012ea} to give numerical estimates of the EOB function $a(u)$ beyond the weak-field (PN) regime $u \ll 1$. In particular, Akcay et al. \cite{Akcay:2012ea} gave accurate numerical representations of the function $a(u)$ over the interval $0 < u < 1/3$, and discovered the presence of a singularity near the ``light-ring'' $u \to 1/3$.

Here, we are interested in the PN regime of $a(u)$, i.e. its expansion in powers of $u$. By inserting in Eq.~(\ref{eq48}) our previous analytic GSF calculation of $h_{kk}^R (u)$, Eq.~(\ref{eq46}), we get the following 4PN-accurate expansion of $a(u)$:
\begin{eqnarray}
\label{eq49}
a(u)&=&  2 u^3+\left(\frac{94}{3}-\frac{41}{32}\pi^2\right) u^4 \nonumber \\
&+&\biggl(-\frac{4237}{60}+\frac{128}{5}\gamma  \\
&&+\frac{64}{5}\ln(u)+\frac{256}{5}\ln(2)+\frac{2275}{512}\pi^2\biggl)u^5 + o(u^5) \, .\nonumber
\end{eqnarray}
This corresponds to the results (\ref{eq5}) and (\ref{eq6b}) given above. 

\subsection{Nonlinear-in-$\nu$ contributions to $A(u;\nu)$ at the 4PN level}


The energetics of comparable-mass binary orbits is fully described, in the EOB formalism, by the function $A(u;\nu)$ \cite{Buonanno:2000ef,Damour:2009sm,Barausse:2011dq,Akcay:2012ea}. More precisely, given the EOB potential $A(u;\nu)$, the total energy $H^{\rm tot} = Mc^2 + E_B$ and the dimensionless frequency parameter $x = (GM\Omega / c^3)^{2/3}$ can be  both computed as explicit functions of $u$ (see, e.g., Section~IV of \cite{Damour:2009sm}). When inserting the PN expansion of $A(u)$ we can get corresponding PN expansions of $ E_B(u)$ and $x(u)$. Inverting the latter expansion (which starts as $x(u) =  u + \frac13 \, \nu u^2 + O(u^3)$) to get $u$ in terms of $x$, we can then straightforwardly obtain the PN expansion of the function relating $ E_B$ to the frequency parameter $x$. It has the form 
\begin{eqnarray}
\label{eq10}
E_B (x;\nu) &= &-\frac12 \mu c^2 x (1+e_{\rm 1PN} (\nu) x + e_{\rm 2PN} (\nu) x^2 \nonumber \\
&+& \, e_{\rm 3PN} (\nu) x^3 + e_{\rm 4PN} (\nu , \ln x) x^4 + o (x^4))
\end{eqnarray}
with a 4PN coefficient (here expressed by using the 2PN result $a_3 (\nu) = 2$, but leaving $a_4 (\nu) = a_4$ and $a_5 (\nu)$ in analytic form)
\begin{eqnarray}
\label{eq58}
e_{\rm 4PN} (\nu,\ln x) &= &-\frac{3969}{128} + \left(\frac{3213}{128} + \frac72 \, a_4 \right) \nu \nonumber \\
&+ &\left( \frac73 \, a_5^c(\nu) + \frac23 \, a_5^{\ln} (\nu) \right) \nu \nonumber \\
&+ &\left( \frac{1015}{384} - \frac{35}{18} \, a_4 \right) \nu^2 + \frac{301}{1728} \, \nu^3 \nonumber \\
&+ &\frac{77}{31104} \, \nu^4 + \frac73 \, a_5^{\ln} (\nu) \, \ln x \, .
\end{eqnarray}


Jaranowski and Sch\"afer \cite{Jaranowski:2013lca,Jaranowski:2012eb} have recently determined the coefficients of 
$\nu^2$, $\nu^3$ and $\nu^4$ in  $e_{\rm 4PN} (\nu,\ln x)  = e_{\rm 4PN}^0  + \nu e_{\rm 4PN}^1 + \nu^2 e_{\rm 4PN}^2 + \nu^3 e_{\rm 4PN}^3  +\frac{448}{15} \nu \ln x $, namely 
\be
\label{eq14}
e_{\rm 4PN}^2 = - \frac{498449}{3456} + \frac{3157}{576} \pi^2 \, ,
\ee
\be
\label{eq15}
e_{\rm 4PN}^3 =  \frac{301}{1728} \, ,
\ee
\be
\label{eq16}
e_{\rm 4PN}^4 =  \frac{77}{31104} \, .
\ee
By comparing Eq.~(\ref{eq58}) to these results we deduce our result   Eq.~(\ref{eq6c}) above.

\section{Discussion}
\label{section:four}

The results presented here complete a line of work which has been started years ago by obtaining the exact, analytic expression of the 4PN contribution to the main potential determining the energetics of circular orbits of comparable-mass binary systems. Our results also open new avenues for further progress. First, the fact that our method has allowed one to unambiguously extract local dynamical information in a situation where formal infrared divergences have recently bogged down a direct calculation of the full interaction Hamiltonian \cite{Jaranowski:2013lca} suggest that it could help to surmount these formal infrared divergences (which our work has clearly related to the old result of Ref.~\cite{Blanchet:1987wq} on the breakdown of the PN scheme). 

Another interesting  avenue opened by our results concerns the nonlinear  dependence in $\nu$ of the EOB radial potential $A(u;\nu)$.   Up to the 3PN level included (i.e. for $n \leq 4$), the contributions $\sim \nu a_n (\nu) u^n$ to $A(u;\nu)$ were {\it linear} in $\nu$, i.e. the coefficient $a_n (\nu)$ was independent of $\nu$. As emphasized in \cite{Buonanno:1998gg,Damour:2000we}, such a linearity in $\nu$ was linked to remarkable cancellations between nonlinear terms in $\nu$ when computing $A(u;\nu)$ from the (Delaunay) Hamiltonian. 
Though similar remarkable cancellations occur at the 4PN level when computing $A(u;\nu)$ from the energy-frequency function $E(\Omega)$ (namely, as emphasized in \cite{Damour:2012mv},  the $O(\nu^3)$ and $O(\nu^4)$ contributions to $E(x;\nu)$  \cite{Jaranowski:2013lca,Jaranowski:2012eb} cancell out when translated in terms of $A(u;\nu)$) such cancellations do not extend to the $O(\nu^2)$ 4PN-level contribution to $A(u;\nu)$.
In particular, our work shows that the $\nu^2 a_2 (u)$ contribution in Eq.~(\ref{eq47}) is {\it negative}, and starts (in the weak-field domain) as
\begin{eqnarray}
\label{eq59}
\nu^2 a_2 (u) &= &\left( -\frac{221}6 + \frac{41}{32} \, \pi^2 \right) \nu^2 u^5 + \nu^2 \, o(u^5) \nonumber \\
&\simeq &-24.1879027 \, \nu^2 u^5 + \nu^2 \, o(u^5) \, .
\end{eqnarray}
The necessity of  a negative $O(\nu^2)$ contribution has been recently suggested (in Sec. VII of \cite{Akcay:2012ea}) and is also apparent in the $\nu$-dependence of the ``effective'' $a(u;\nu)$ functions obtained by comparing the EOB formalism to accurate numerical relativity simulations of binary black holes (see, in particular, Fig.~16 in \cite{Damour:2012ky}). This suggests that one should include our $\nu$-dependent value of $a_5^c (\nu)$ within the Pad\'e-resummed expressions used to parametrize the $A(u;\nu)$ potential that is compared to numerical relativity simulations. 

As for the numerical value of the $\nu \to 0$ limit of $a_5^c (\nu)$, our analytic result (\ref{eq6b}) yields
\be
\label{eq8}
a_{5 \, {\rm ana}}^{c0} = 23 . 5033892426 \ldots
\ee
Recently, accurate numerical computations of Detweiler's redshift function \cite{Blanchet:2009sd,Blanchet:2010zd} have been used (see \cite{Tiec:2011ab, Tiec:2011dp,Barausse:2011dq}) to infer the following numerical estimate of $a_5^{c0}$,
\be
\label{eq9}
a_{5 \, {\rm num}}^{c0} = 23 . 50190(5) \, .
\ee
We note that the first four digits of our analytic result (\ref{eq8}) nicely agree with those of the previous numerical estimate (\ref{eq9}). However, the two results differ by $a_{5 \, {\rm num}}^{c0} - a_{5 \, {\rm ana}}^{c0} \simeq - 0.00149(5)$, which is $30$ times larger than the estimated error bar on the numerical value of $a_5^{c0}$. We think that this difference is due to an optimistic view of the numerical accuracy on the determination of $a_5^{c0}$. Indeed, though the value (\ref{eq9}) corresponds to the ``best fit'' estimate of the related 4PN coefficient, say $a_4^{\rm BDLW}$ (which is denoted $a_4$ in \cite{Blanchet:2010zd}), the scatter among the various fitted values of $a_4^{\rm BDLW}$ summarized in Table~VI of \cite{Blanchet:2010zd} is compatible with our analytic estimate (\ref{eq8}). Indeed, the latter total scatter (from the difference Table II $-$ Table III),  $\delta a_4^{\rm BDLW} \simeq 0.0027$, translates into a scatter in $a_5^0$ of order $\delta a_5^{c0} = \frac12 a_4^{\rm BDLW} \simeq 0.0014$ which is comparable to the difference $\vert a_{5 \, {\rm num}}^{c0} - a_{5 \, {\rm ana}}^{c0} \vert$. This highlights the need to be conservative when estimating uncertainties on parameters obtained from fitting numerical data.
Let us also note that the analytical value of the $\nu-$linear coefficient  $e_{\rm 4PN}^1$ in the 4PN term in the energy-frequency function $E_B(x)$ reads
\begin{eqnarray}
\label{eq13}
e_{\rm 4PN}^1 &= &\frac{91713}{640} - \frac{287}{64} \pi^2 + \frac73 a_5^{c0} \nonumber \\
&= &-\frac{123671}{5760} + \frac{9037}{1536} \pi^2 + \frac{1792}{15} \ln 2 + \frac{896}{15} \gamma \, .
\end{eqnarray}
The corresponding exact numerical value is
\be
\label{eq17}
e_{\rm 4PN \, ana}^1 = 153 . 8837968 \ldots 
\ee
This differs by $+ \, 0.0035(1)$ from the value $e_{\rm 4PN \, num}^1 \simeq 153.8803(1)$ estimated in \cite{Tiec:2011ab} from the numerically-fitted $a_4^{\rm BDLW}$ of \cite{Blanchet:2010zd}. Again, this difference is comparable to the one induced by the above-quoted scatter among the various fitted values of $a_4^{\rm BDLW}$. [Note that $\delta e_{\rm 4PN}^1 = \frac73 \delta a_5^{c0} = \frac76 \delta a_4^{\rm BDLW}$.]

Finally, let us mention that the $\nu$-linear GSF part of our work is not limited to the 4PN level, and that we intend to extend it to higher PN accuracies.

\bigskip

\noindent {\bf Acknowledgments.} T.D. thanks Pierre Deligne for informative discussions. We are grateful to ICRANet for partial support. D.B. thanks IHES for hospitality during crucial stages of development of this project.


\end{document}